\documentclass{emulateapj}
\usepackage{psfig}
\usepackage{apjfonts,times}

\newcommand{\beq}{\begin{equation}}
\newcommand{\eeq}{\end{equation}}

\def\alp{\mbox{$\alpha$}}

\def\farcs{\hbox{$.\!\!^{\prime\prime}$}}
\def\earth{\hbox{$\oplus$}}
\def\arcmin{\hbox{$^\prime$}}

\def\solar{\mbox{$_{\normalsize\odot}$}}

\newcommand{\AmS}{{\protect\the\textfont2
  A\kern-.1667em\lower.5ex\hbox{M}\kern-.125emS}}
\newcommand{\lsim}{\ \raise
-2.truept\hbox{\rlap{\hbox{$\sim$}}\raise5.truept\hbox{$<$}\ }}
\newcommand{\gsim}{\ \raise
-2.truept\hbox{\rlap{\hbox{$\sim$}}\raise5.truept\hbox{$>$}\ }}
\newcommand{\simsim}{\ \raise
-2.truept\hbox{\rlap{\hbox{$\sim$}}\raise5.truept\hbox{$\sim$}\ }}

\slugcomment{Accepted for publication in The Astrophysical Journal 
Supplement Series, 22 June 2006}

\shorttitle{HST/ACS Photometry of the Star-Forming Region NGC 346 in the SMC}
\shortauthors{D. A. Gouliermis et al.}

\begin{document}

\title{The Star-Forming Region NGC 346 in the Small 
Magellanic Cloud with Hubble Space Telescope ACS Observations I. 
Photometry\footnotemark[1]}
\footnotetext[1]{Research supported by the Deutsche 
Forschungsgemeinschaft (German Research Foundation)}

\author{D. A. Gouliermis\altaffilmark{2}, 
        A. E. Dolphin\altaffilmark{3}, 
        W. Brandner\altaffilmark{2}, 
        Th. Henning\altaffilmark{2}}
\altaffiltext{2}{Max-Planck-Institut f\"{u}r Astronomie, K\"{o}nigstuhl 
17, D-69117 Heidelberg, Germany, dgoulier@mpia.de, brandner@mpia.de, 
henning@mpia.de}
\altaffiltext{3}{Steward Observatory, 933 N. Cherry Avenue Tucson, AZ 
85721-0065, USA, adolphin@as.arizona.edu}

\begin{abstract}

We present a photometric study of the star-forming region NGC 346 and its 
surrounding field in the Small Magellanic Cloud, using data taken with the 
Advanced Camera for Surveys (ACS) on board the Hubble Space Telescope 
(HST).  The data set contains both short and long exposures for increased 
dynamic range, and photometry was performed using the ACS module of the 
stellar photometry package DOLPHOT.  We detected almost 100,000 stars over 
a magnitude range of $V \sim 11$ to $V \sim 28$ mag, including all stellar 
types from the most massive young stars to faint lower main sequence and 
pre-main sequence stars. We find that this region, which is characterized 
by a plethora of stellar systems and interesting objects, is an outstanding 
example of mixed stellar populations. We take into account different 
features of the color-magnitude diagram of all the detected stars to 
distinguish the two dominant stellar systems: The stellar association NGC 
346 and the old spherical star cluster BS 90. These observations provide a 
complete stellar sample of a field about 5\arcmin\ $\times$ 5\arcmin\ 
around the most active star-forming region in this galaxy. Considering the 
importance of these data for various investigations in the area, we 
provide the full stellar catalog from our photometry. This paper is the 
first part of an ongoing study to investigate in detail the two dominant 
stellar systems in the area and their surrounding field.

\end{abstract}


\begin{figure*}[t!]
\epsscale{1}
\plotone{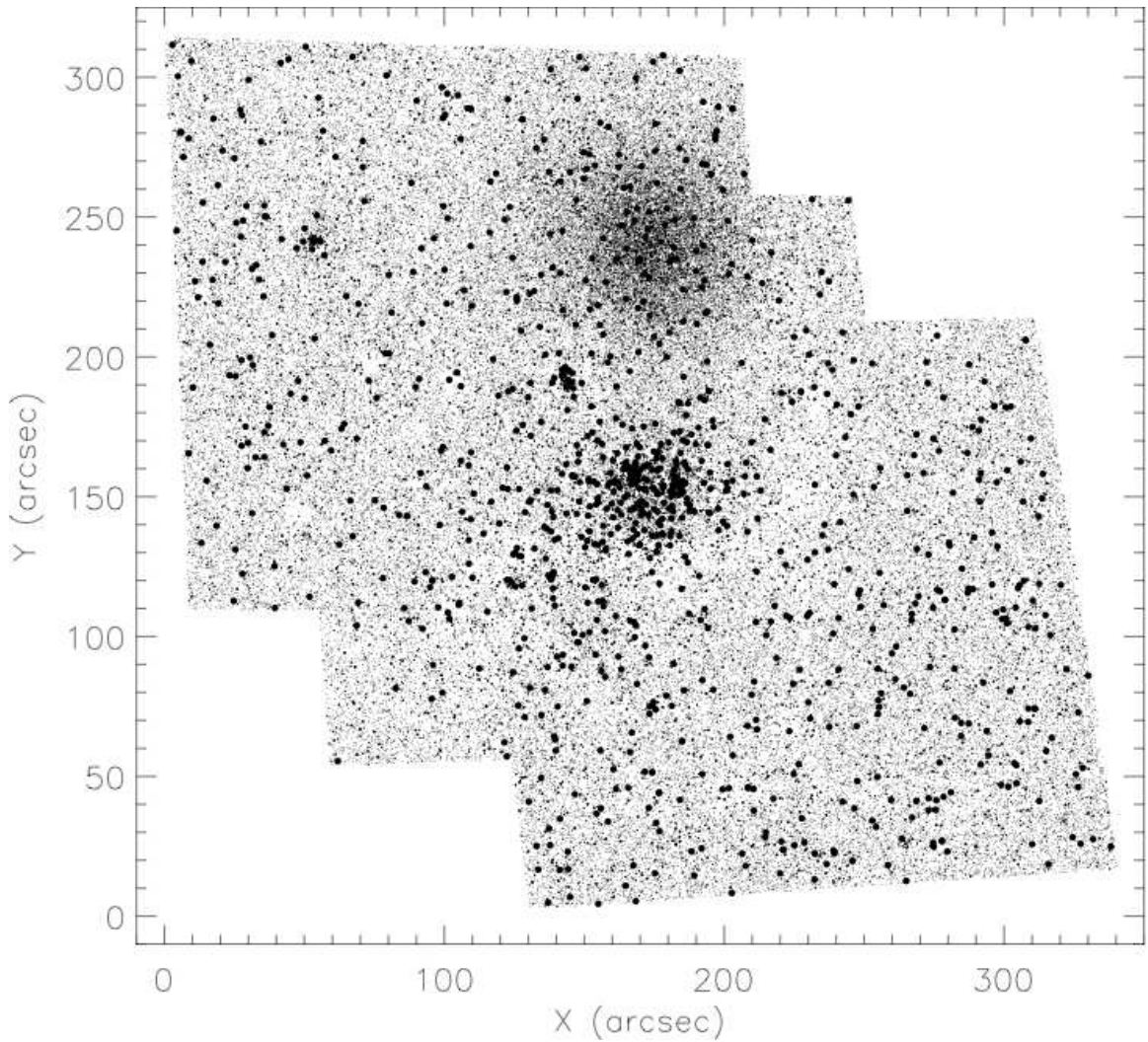}
\caption{Map of stars found in both F555W and F814W filters with DOLPHOT
photometry based on ACS/WFC imaging of three fields. The map covers the
whole area around the association NGC 346 in the SMC. North is up and
east is left. Coordinates refer to the drizzled F814W image and are
given in seconds of arc.} 
\label{fig-map} 
\end{figure*}

\section{ Introduction}

The investigation we present here deals with the stellar content of the
star forming region in the vicinity of the stellar association NGC 346
in the Small Magellanic Cloud (SMC). NGC 346 is the largest stellar
concentration in the SMC, related to the brightest {\sc Hii} region of
this galaxy, named LHA 115-N 66 or in short N 66 (Henize 1956), with an
H{\alp} luminosity almost 60 times higher than Orion's (Kennicutt 1984).
Such active star forming regions are rare and can only be compared to
star-bursts, like e.g., NGC 3603 in our Galaxy (Sher 1964).  As a
consequence NGC 346 has been the subject of intense studies to establish
the properties of the {\sc Hii} region and of the bright early-type
stellar content of the association in the low-metallicity environment of
the SMC.

In a spectroscopic study of 42 bright main sequence stars Massey et al. 
(1989) confirmed 33 O-stars in the area of NGC 346. 22 of them are located 
in the central region of the association, and eleven are of type O6.5 or 
earlier. The stellar sample of Massey et al. is essentially complete down 
to $\sim$ 10 M{\solar} with six stars in the mass range 40 - 85 M{\solar} 
and with a background field population of $\sim$ 5 M{\solar} stars.

An ultraviolet and optical spectral atlas of O stars in the SMC is 
compiled by Walborn et al. (2000) with data from the HST Space Telescope 
Imaging Spectrograph (STIS), AAT and ESO 3.6 m. Systematic phenomena in 
the sample include weak stellar-wind profiles on the O main sequence and 
in late-O giants, the absence of Si {\sc iv} wind features throughout the 
O-giant sequence, low terminal velocities and enhanced He {\sc ii} wind 
features in Of super-giants. The spectra of six O-type stars in NGC 346 
from the Walborn et al. (2000) sample were modeled by Bouret et al. (2003) 
to determine their chemical abundances and wind parameters. The majority 
of the stars reveal CNO cycle-processed material at their surfaces during 
the main-sequence stage, thus indicating fast stellar rotation and/or very 
efficient mixing processes. The derived effective temperatures are lower 
than predicted from the widely used relation between spectral type and 
$T_{\rm eff}$, resulting in lower stellar luminosities and lower ionizing 
fluxes.

However, Massey et al. (2005) argue that the data used by Bouret et al. 
(2003) were limited with respect to sky (nebular) subtraction. The 
effective temperatures of O3 - O7 dwarfs and giants in the SMC, as found 
by Massey et al. (2005) from their investigation of 33 O-type stars in the 
Magellanic Clouds, are about 4000 K hotter than for stars of the same 
spectral type in the Milky Way. Consequently, the effective temperature 
scale for O dwarfs in NGC 346 derived by these authors is significantly 
hotter than the Bouret et al. (2003) values. The differences decrease as 
one approaches stars of spectral type B0~V. In general, Massey et al. 
(2005) found that the winds momentum of O stars in the Magellanic Clouds 
scales with luminosity and metallicity as predicted by the theory for 
radiatively driven wind (Kudritzki 2002; Vink et al. 2001), supporting the 
use of photospheric analyses of hot luminous stars as a distance indicator 
for galaxies with resolved massive stellar populations.

A search for Be-stars in a field of 10\arcmin\ $\times$ 10\arcmin\ 
centered on NGC 346 (Keller et al. 1999) found that only about 11\% of the 
upper main sequence stars (with V $<$ 16 mag) located in the main body of 
the association show indeed strong H\alp\ emission. These authors adopt 
the most widely accepted explanation for the Be phenomenon, the rapid 
rotation of the B star (Bjorkman \& Cassinelli 1993), due to which the 
stellar wind from a star is concentrated and confined to a disk around its 
equatorial plane. They relate, thus, the low fraction of Be stars in the 
cluster with the suggestion that it has been formed from low angular 
momentum gas.

The general area of NGC 346/N 66 hosts at least two supernova remnants.  
Not much is known about the first, named SNR 0056$-$7233, which has been 
identified by Ye et al. (1991) to the southwest at a projected distance of 
about 5.4 pc from the center of NGC 346. The second supernova remnant, 
named SNR 0057$-$7226, is located northeast close to the Wolf-Rayet 
luminous blue variable system HD 5980 in the vicinity of NGC 346.  
Koenigsberger et al. (2001) observed with HST/STIS interstellar and 
circumstellar absorption components along the line of sight toward HD 
5980. They argued that SNR 0057$-$7226 had probably a progenitor, that was 
one of the brightest members of the association. An estimation of $T \sim$ 
40,000 K and a total mass between 400 and 1000 M{\solar} for the supernova 
remnant shell was made by these authors. \textit{Chandra} observations 
have shown that HD 5980 is surrounded by a region of diffuse X-ray 
emission from the supernova remnant, while NGC 346 itself shows only 
relatively faint X-ray emission, most of which seems correlated with the 
location of the brightest stars in the core of the association (Naz{\'e} 
et al. 2003).

\begin{figure*}[t!]
\centerline{\hbox{
\includegraphics[width=8.75truecm,angle=0]{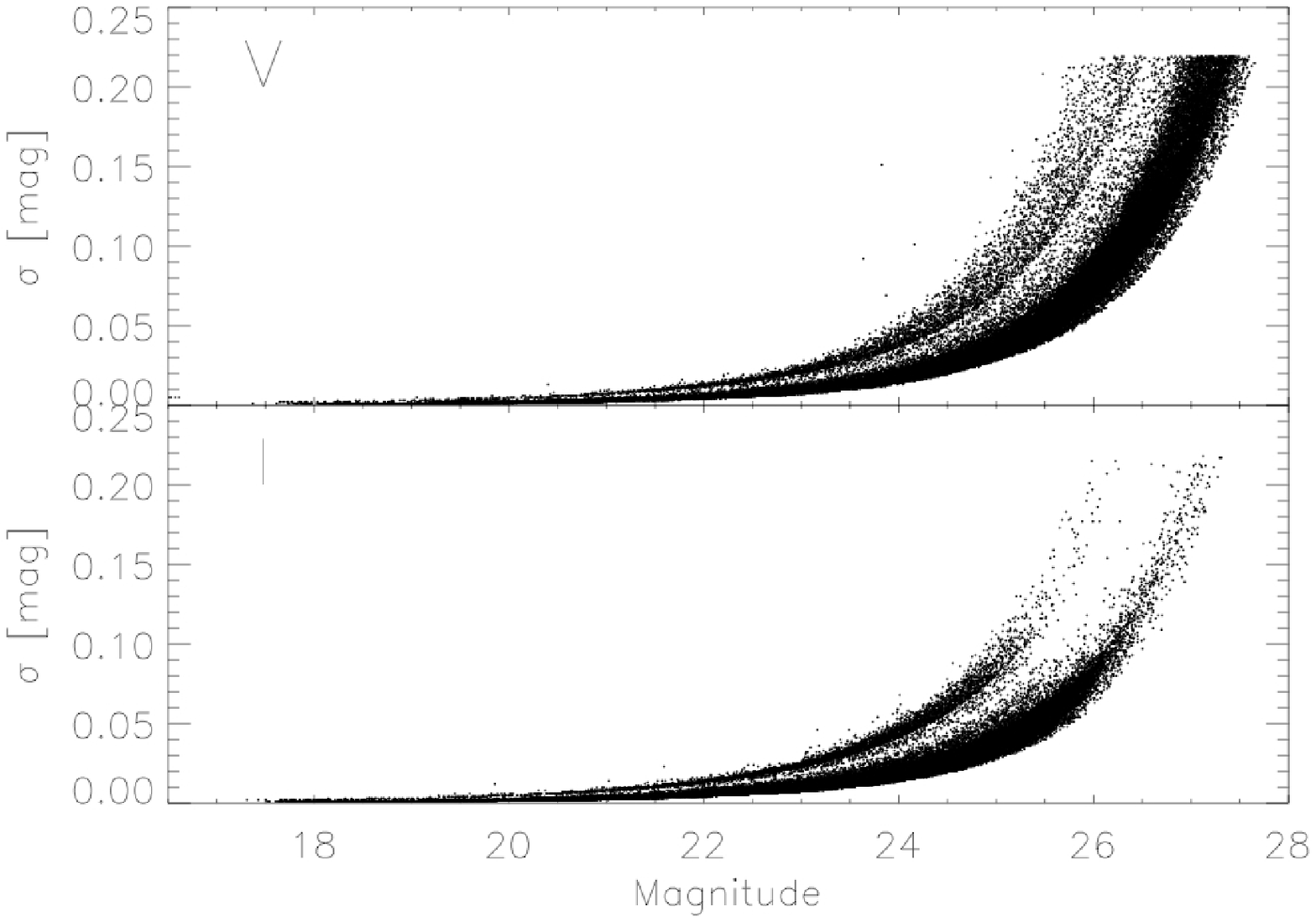}
\includegraphics[width=9.truecm,angle=0]{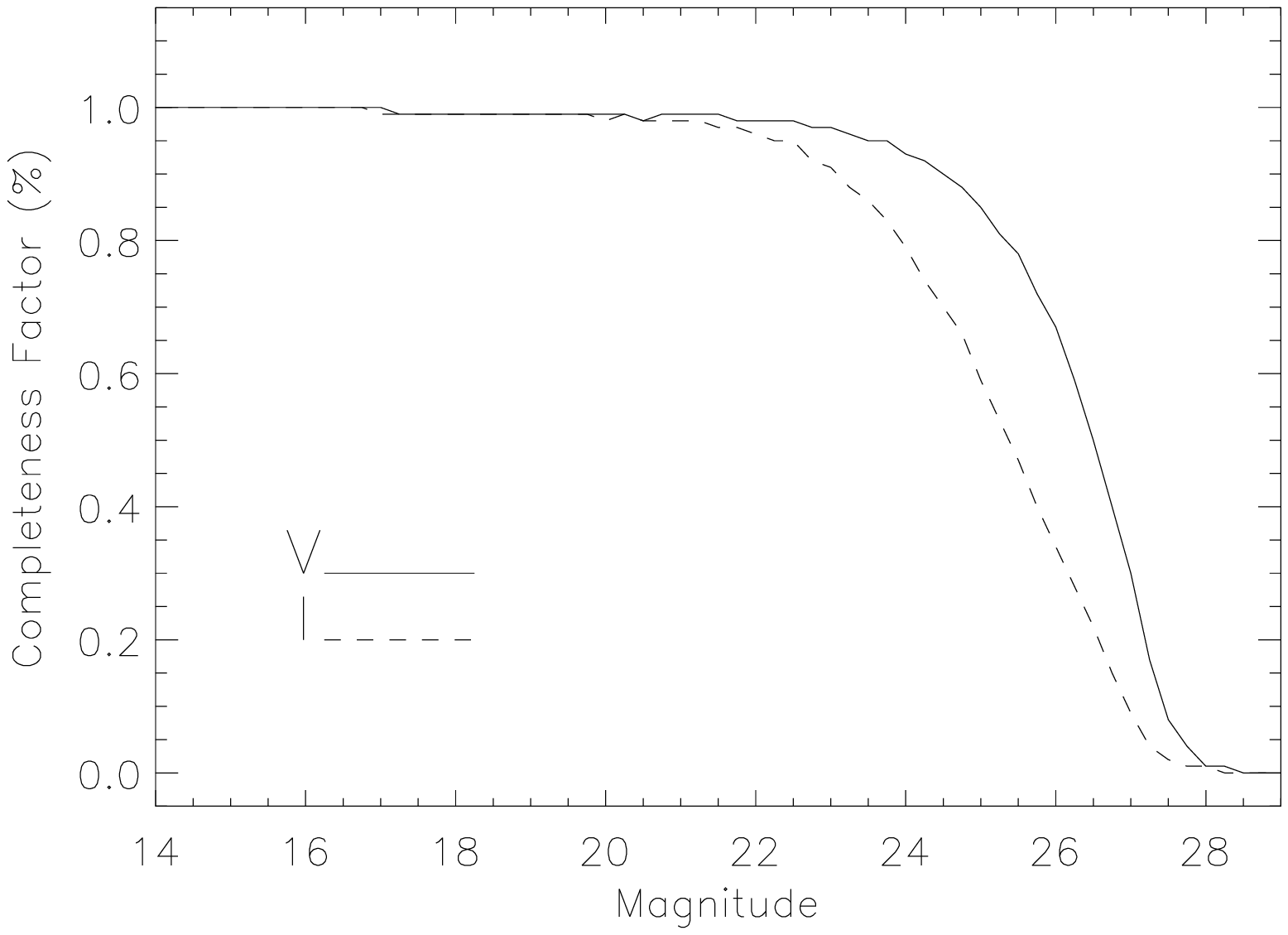}
}}
\caption{Typical uncertainties of photometry (left) and completeness 
factors (right) as derived by DOLPHOT from all datasets, for both bands. 
The bifurcation in the photometric errors is due to varying depth within
the overall field.  The stars in the high-sigma branch are located in
regions covered only by single exposures (dataset J92FA3).}
\label{fig_pherr-cmp}
\end{figure*}

Danforth et al. (2003) found with the {\em Far Ultraviolet Spectroscopic 
Explorer} (FUSE) strong {\sc O vi} and {\sc C iii} emission from a 
position at the edge of SNR 0057$-$7226. They determined the physical 
parameters of the interaction zone with N 66 and found that ionizing 
photons from massive stars of the association likely affect the ionization 
balance in the post-shock gas, hindering the production of lower 
ionization and neutral species. Danforth et al. (2003) proposed a physical 
relationship between SNR 0057$-$7226 and N 66, which suggests a slowly 
expanding bubble around NGC 346, powered by stellar winds and thermal 
pressure, while strong {\sc O vi}, {\sc C iii}, and X-ray emission arise 
from the shock interaction of the super-nova remnant with the denser, 
ionized gas on the rear side of N 66.

Rela{\~n}o et al. (2002) computed photoionization models of the {\sc H ii} 
region around NGC 346 with the CLOUDY 94 algorithm (Ferland et al. 1998) 
based on spectrophotometric data of Peimbert et al. (2000). These models 
showed that about 45\% of the photons produced by the ionizing stars 
escape from NGC 346, implying that the system is a major source of 
ionizing flux for the surrounding diffuse interstellar medium. They also 
suggested that N 66 is a spherical density-bounded nebula.

The recent release of the {\em Advanced Camera for Surveys} imaging
toward the general area of NGC 346 from the {\em HST Data Archive}
offers a unique opportunity for the detailed photometric study not only
of the bright massive stellar content, which is well known, but mostly
of the fainter stellar members of the area down to the lower main
sequence stars, which represent the majority of the stellar
populations. In this paper we present our photometry of these 
observations, that cover a field about 5\arcmin\ $\times$ 5\arcmin 
centered on the association NGC 346. Our photometry detects magnitudes 
as faint as $V$ \gsim\ 26.5 mag, providing one of the richest 
stellar samples observed in the SMC. We also describe the stellar content 
of the area as it is seen in the derived color-magnitude diagram. 

Specifically, in section 2 we present the observations and the photometry 
with software especially designed for ACS imaging. In the following 
section we present the stellar catalog and the color-magnitude diagram. We 
identify the different stellar populations coexisting in the area and we 
discuss their spatial distribution (\S 3). A general description of the 
stellar content of the two main stellar concentrations in the area, the 
young stellar association NGC 346, and the old spherical cluster [BS95] 90 
or in short BS 90 (Bica \& Schmitt 1995) is given in section 4.  Final 
remarks and future prospects of the science that this catalog can provide 
are discussed in section 5.


\begin{deluxetable}{ccccc}
\tablewidth{0pt}
\tablecaption{Log of the observations. Datasets refer to HST archive 
catalog. Exposure times (T$_{\rm expo}$) per filter are given in seconds. 
\label{tab1}}
\tablehead{\colhead{Visit} & \colhead{RA} & \colhead{DEC} & 
\multicolumn{2}{c}{T$_{\rm expo}$}\\
\colhead{Dataset} & \multicolumn{2}{c}{(J2000)}& \colhead{F555W} & 
\colhead{F814W}}
\startdata
J92F01&00:59:07.61&$-$72:09:39.8&1$\times$456&1$\times$484\\
      &00:59:07.31&$-$72:09:40.0&1$\times$456&1$\times$484\\
      &00:59:07.02&$-$72:09:40.1&1$\times$456&1$\times$484\\
      &00:59:06.72&$-$72:09:40.3&1$\times$456&1$\times$484\\
      &00:59:06.72&$-$72:09:40.3&2$\times$3&2$\times$2\\
\tableline
J92F02&00:59:05.58&$-$72:11:21.5&1$\times$483&1$\times$450\\
      &00:59:05.28&$-$72:11:21.6&1$\times$483&1$\times$450\\
      &00:59:04.99&$-$72:11:21.6&1$\times$483&1$\times$450\\
      &00:59:04.68&$-$72:11:21.7&1$\times$483&1$\times$450\\
      &00:59:04.68&$-$72:11:21.7&2$\times$3&2$\times$2\\
\tableline
J92FA3&00:59:06.18&$-$72:10:31.2&1$\times$380&1$\times$380\\
\enddata
\end{deluxetable}

\section{ Observations and Photometry}

The data were collected within the HST GO Program 10248 (PI A. Nota). 
Three pointings (with significant offsets) within a single field centered 
on the association NGC 346 were observed with the Wide-Field Channel (WFC) 
of ACS in the filters F555W and F814W, which are equivalent to standard $V$ 
and $I$ bands respectively. We obtained the data from the HST Data 
Archive. A detailed description of the datasets can be found in Table 1. 
The first visit (J92F01) covers the northwestern part of the observed 
region, while the second visit (J92F02) covers the southeastern part.  
Each visit consists of long F555W and F814W exposures taken in a 4-point 
dither pattern, plus two short exposures in each filter at one pointing.  
There is also a third, central visit (J92FA3), which includes a single 
exposure in each filter (as well as H{\alp} and HRC images, which we are 
not using). The whole set of data provides a unique collection of deep 
observations toward the SMC. In addition, the short exposures complement 
the stellar sample with the more massive stars of the area, which occupy 
both the main sequence and the red giant branch as it will be shown later.

We obtained the pipeline-reduced FITS files from the HST Data Archive. In 
order to clean the images of residual warm pixels and cosmic rays, we ran 
\textit{multidrizzle} (Koekemoer et al. 2002) using the recommended recipe 
for multiple visits from the ACS data handbook.  The process involved 
first obtaining separately drizzled images, measuring the alignment 
offsets, and then creating the final drizzled image.  Although we avoid 
photometry on drizzled frames whenever possible (due to position-dependent 
smoothing and PSF blurring), the drizzling process is useful for 
generating a deep reference image of the entire field, and finding cosmic 
rays and bad pixels in the original images.

\begin{deluxetable}{rcccccc}
\tablewidth{0pt}
\tablecaption{Sample from the photometric catalog of stars found in 
this study in the region of NGC 346 with HST/ACS imaging. Magnitudes 
are given in the Vega system.\label{tab2}}
\tablehead{
\colhead{Star} & \colhead{RA} & \colhead{DEC} &
\colhead{V} & \colhead{$\sigma_{V}$} & \colhead{I} &
\colhead{$\sigma_{I}$}\\
\colhead{No.} & \multicolumn{2}{c}{(J2000)} & \multicolumn{2}{c}{(mag)}
& \multicolumn{2}{c}{(mag)}
}
\startdata
      1& 00:58:42.42& $-$72:09:43.02&  12.171&   0.001&  11.491&   0.001\\
      2& 00:59:04.46& $-$72:10:24.49&  12.443&   0.001&  12.705&   0.001\\
      3& 00:58:53.89& $-$72:12:04.50&  13.526&   0.002&  11.887&   0.001\\
      4& 00:58:36.09& $-$72:11:13.31&  13.783&   0.002&  12.080&   0.001\\
      5& 00:59:00.72& $-$72:10:27.91&  13.519&   0.001&  13.750&   0.002\\
      6& 00:59:00.01& $-$72:10:37.67&  13.692&   0.001&  13.923&   0.002\\
      7& 00:59:36.51& $-$72:10:22.33&  14.382&   0.003&  14.386&   0.004\\
      8& 00:58:36.18& $-$72:12:16.56&  14.503&   0.003&  12.958&   0.002\\
      9& 00:58:57.36& $-$72:10:33.38&  13.977&   0.002&  14.193&   0.002\\
     10& 00:59:01.77& $-$72:10:30.94&  14.375&   0.002&  14.466&   0.004\\
     11& 00:59:30.34& $-$72:09:09.40&  14.578&   0.003&  14.721&   0.004\\
     12& 00:59:41.39& $-$72:08:10.14&  14.633&   0.003&  14.760&   0.004\\
     13& 00:59:06.71& $-$72:10:41.02&  14.654&   0.002&  14.812&   0.003\\
     14& 00:59:06.31& $-$72:09:55.84&  14.374&   0.002&  14.365&   0.003\\
     15& 00:59:02.88& $-$72:10:34.68&  14.448&   0.002&  14.700&   0.003\\
\enddata
\label{tab2} 
\end{deluxetable}

Photometry was obtained using the ACS module of the package DOLPHOT 
(version 1.0; Dolphin, in preparation). The short images could not be 
combined in the drizzle process in an easy way. These images were cleaned 
by the \textit{imcombine} task of DOLPHOT.  This was possible because the 
short images were taken in pairs with identical pointings, and thus the 
CRREJ-like cleaning algorithm was sufficient to remove image defects. All 
22 exposures were photometered simultaneously, using the F814W drizzled 
frame as the position reference.  All photometry parameters were set equal 
to the recommended values in the DOLPHOT manual\footnote{DOLPHOT, 
including the ACS module and documentation, can be found at the web-site 
{\tt http://purcell.as.arizona.edu/dolphot/}}. Photometric calibrations 
and transformations were made according to Sirianni et al. (2005), and CTE 
corrections were made according to ACS ISR 04-06.

In total we had 149,253 detections with $S/N \ge 5$ in both filters and 
object classifications as stellar.  To clean bad detections from our 
photometric catalog, we applied two cuts using DOLPHOT's star quality 
parameters.  First, to avoid residuals near bright stars, we eliminated 
all detections with a mean crowding parameter of 0.75 or higher (those 
objects that would be measured twice as bright if nearby stars were not 
subtracted).  Second, to eliminate other non-stellar features, we 
eliminated all objects whose mean sharpness was less than -0.15 (generally 
cosmic rays) or greater than 0.15 (extended sources). After this procedure 
more than 99,000 stars were detected with our photometry (Figure 
\ref{fig-map}). The full length of the photometric catalog with the X, Y 
positions and the V- and I-equivalent magnitudes (in the Vega system) of 
these stars is available in electronic form. Figure \ref{fig_pherr-cmp} 
(left panel) shows typical uncertainties of photometry as a function of 
the magnitude for both filters. The completeness of the data was evaluated 
by artificial star experiments, that were performed by running DOLPHOT in 
artificial star mode with the use of artificial star lists created with 
the utility {\em acsfakelist} of the ACS module for DOLPHOT.  The 
completeness is found to be spatially variable, depending on the crowding 
of each region. The completeness functions for the whole observed field 
are shown in Figure \ref{fig_pherr-cmp} (right panel) for both filters.

\begin{figure}[t!]
\epsscale{1.}
\plotone{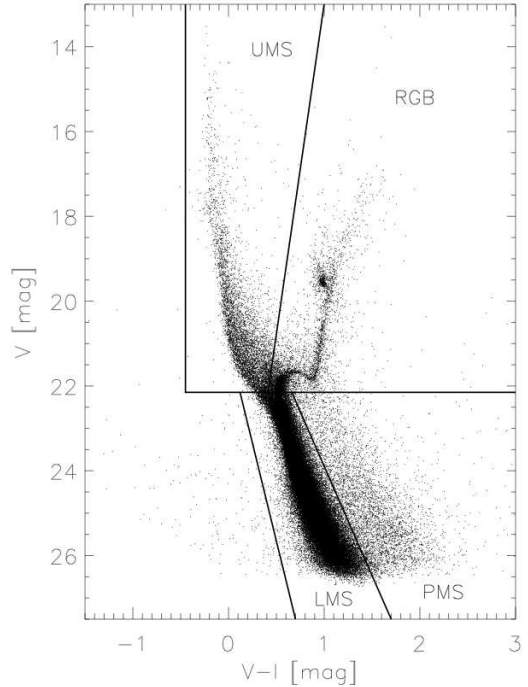}
\caption{$V-I$, $V$ Color-Magnitude Diagram (CMD) of all stars 
detected with the best photometry (see text) in the area of NGC 346 with 
ACS/WFC imaging. The limits of four regions, indicative of specific 
stellar populations, are overplotted. These regions represent the upper 
main sequence (UMS), the lower main sequence (LMS), the red giant branch 
(RGB) and the region where candidate pre-main sequence (PMS) stars have 
been previously identified.}
\label{fig-cmd} 
\end{figure}

\begin{figure*}[t!]
\centerline{\hbox{
\includegraphics[width=0.450\textwidth,angle=0]{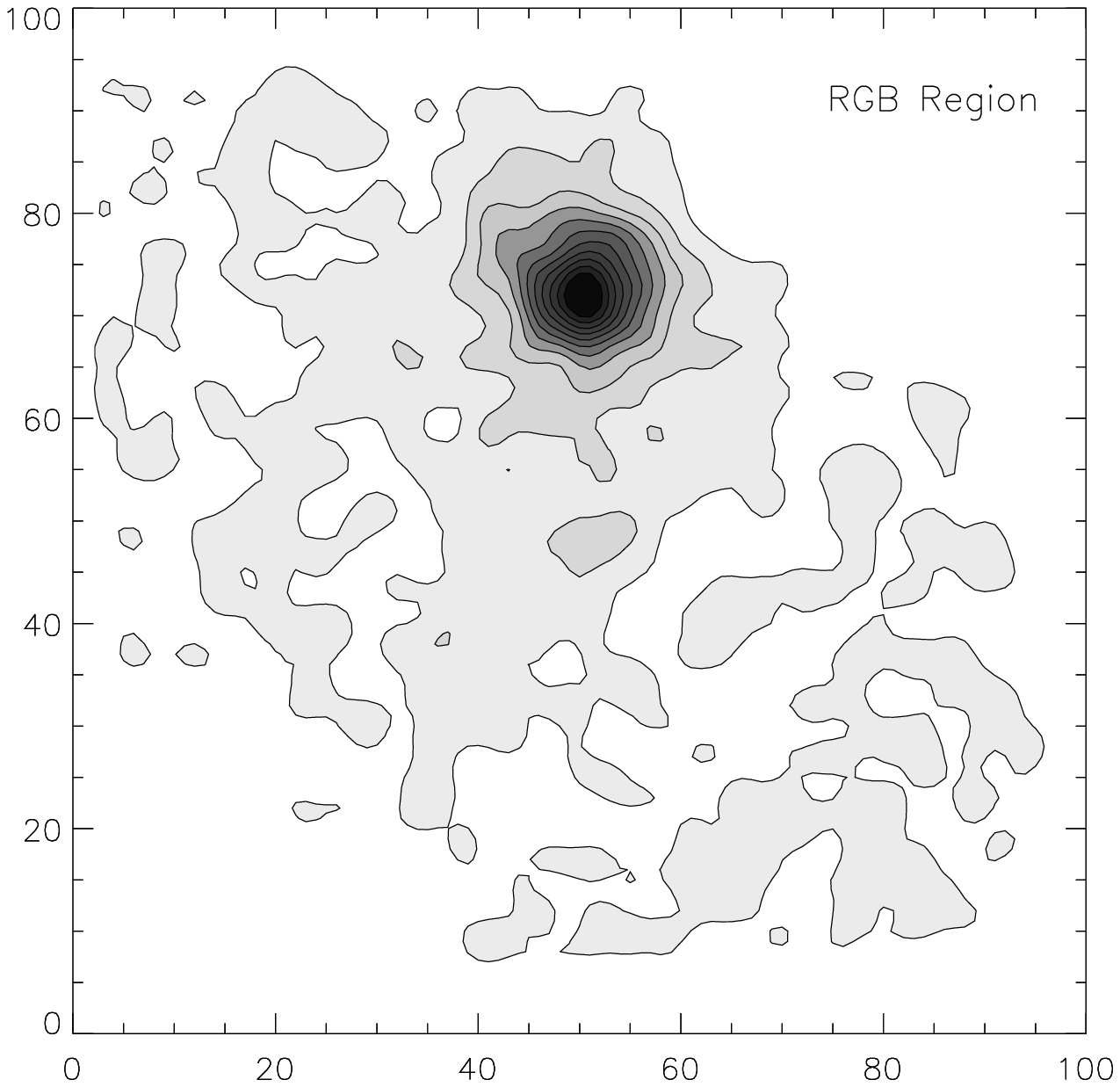}
\includegraphics[width=0.450\textwidth,angle=0]{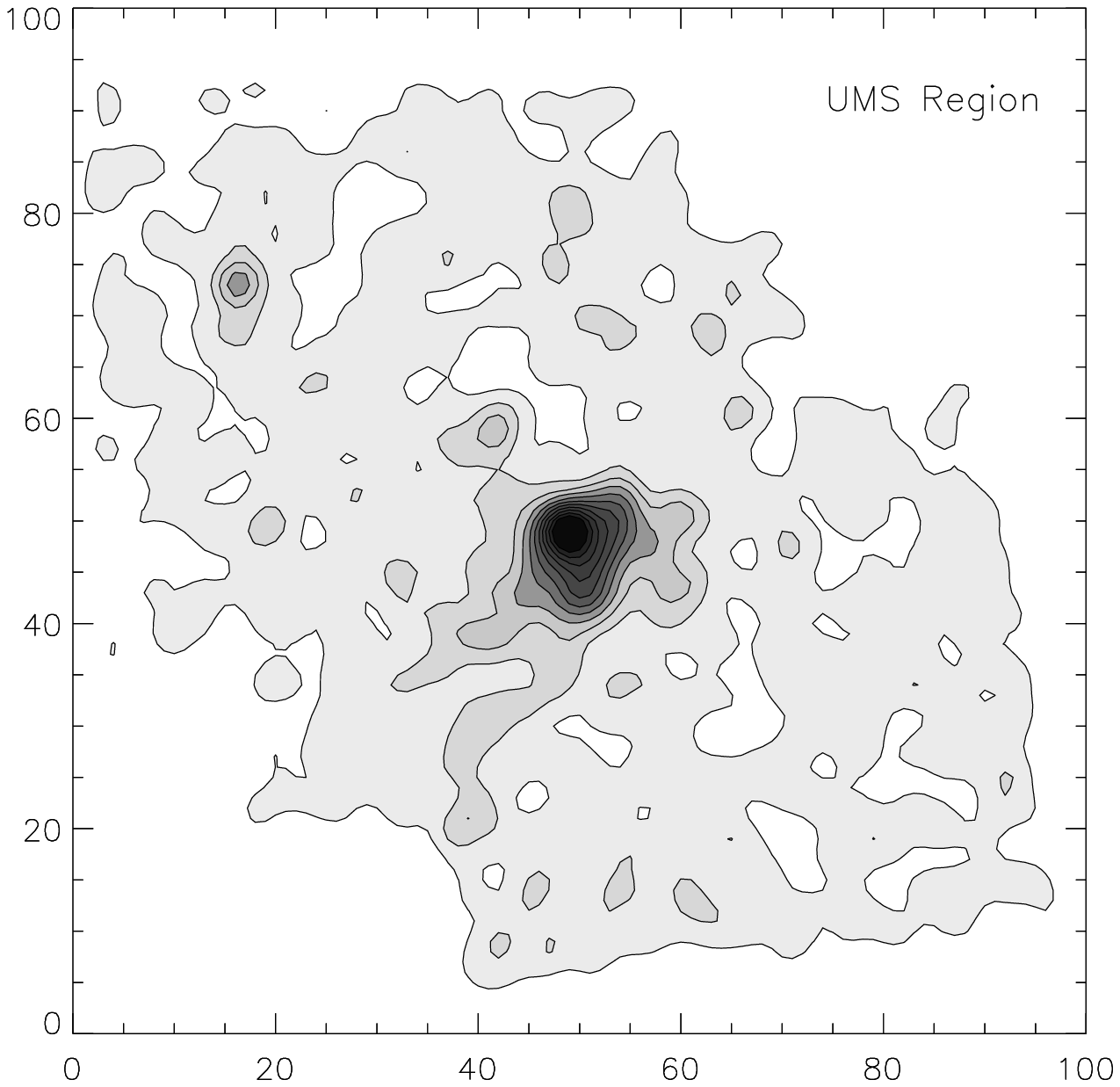}
}}
\centerline{\hbox{
\includegraphics[width=0.450\textwidth,angle=0]{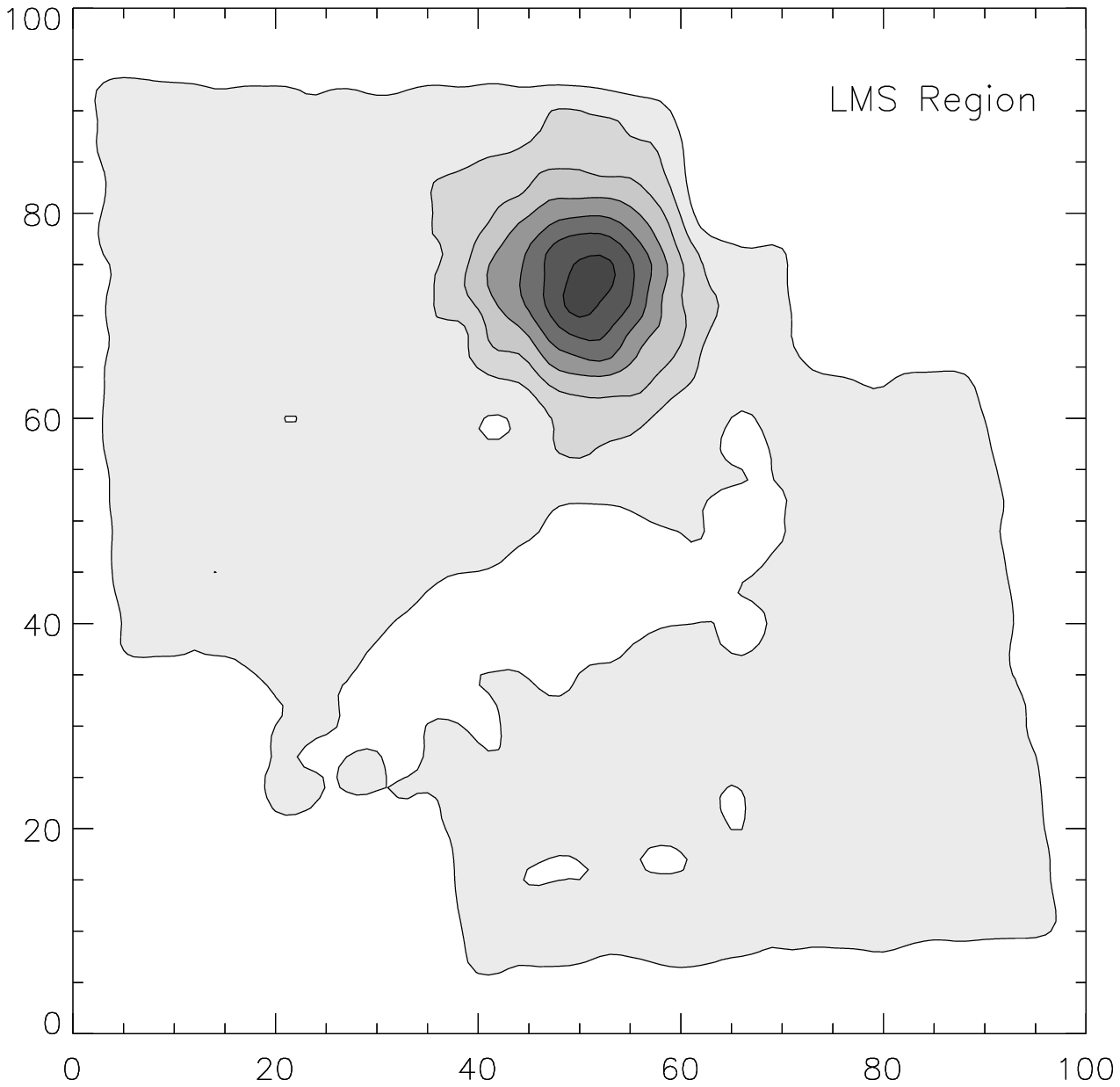}
\includegraphics[width=0.450\textwidth,angle=0]{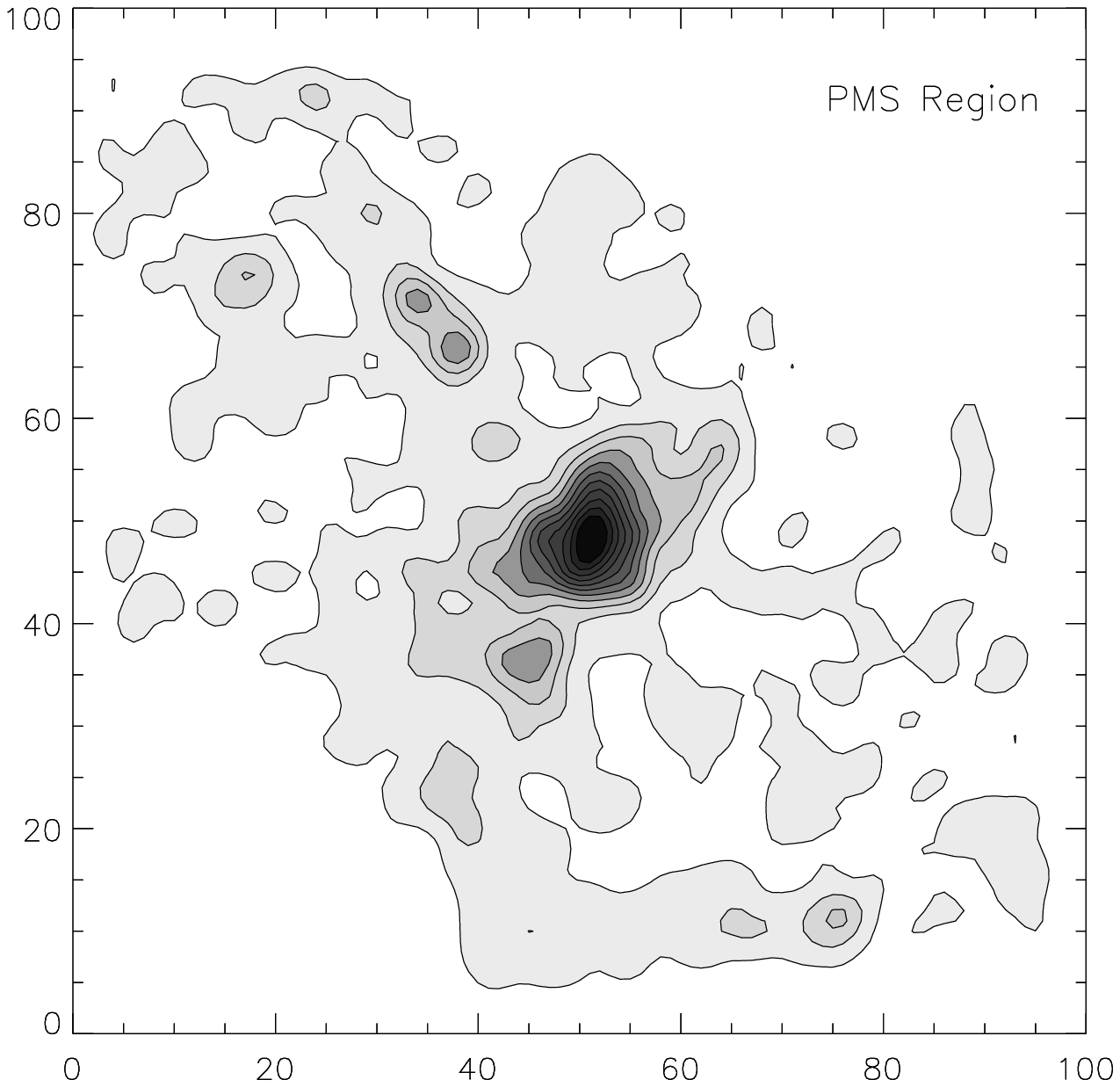}
}}
\caption{Isodensity contour maps constructed with star counts of four 
different stellar populations, which cover different regions in the CMD of 
Figure \ref{fig-cmd} (RGB, UMS, LMS, and PMS).}
\label{fig-con}
\end{figure*}

\section{Stellar Populations}

\subsection{The Color-Magnitude Diagram}

The region of NGC 346 is found to provide an excellent example of well 
resolved extra-galactic stellar populations. The $V-I$, $V$ 
Color-Magnitude Diagram (CMD) of the detected stars (Figure \ref{fig-cmd}) 
shows that the region is characterized by the coexistence of 
several stellar populations. There is a sharp upper main sequence (UMS) of 
young blue massive stars, a clear turn-off at around $V\simeq 22$ mag and 
a prominent red giant branch (RGB) with its red clump clearly located at 
around $V \simeq 19$ mag and $V-I \simeq 1.0$ mag. Below the turn-off the 
main sequence is becoming highly populated with what should be considered 
to be a well mixed collection of low-mass stars of several ages. Under 
this assumption, the lower main sequence (LMS) represents stars, which are 
located in one region of the CMD, but they could have been born during 
completely different star formation events. To the right of LMS and toward 
redder colors we find a concentration of low-mass stars, that do not fit 
to any classical scheme of stellar populations. Most likely they are 
pre-main sequence (PMS) stars.

Such stars have been found to exist in stellar associations of the
Magellanic Clouds. Gouliermis et al. (2006) presented recently the case
of the association LH 52 in the Large Magellanic Cloud, where they
identified the candidate PMS population of the system with HST/WFPC2
observations in $V$ and $I$. The detected PMS stars of LH 52 have the
same location on the $V-I$, $V$ CMD with the faint red sequence we
observe in the CMD of Figure \ref{fig-cmd} to the right of the LMS. As
far as the association NGC 346 is concerned, Brandner et al. (1999)
already reported the detection of about 150 objects with excess emission
in H{\alp} in a 2\arcmin\ $\times$ 2\arcmin\ field slightly off the
brightest stars of the association. They explained this detection as an
indication that NGC 346 hosts PMS stars with masses between 1 and 2
M{\solar}. More recently, Nota et al. (2006) used their observations of
NGC 346, taken within Visit J92FA3 (see table 1) in filters F555W ($V$),
F814W ($I$) and F658N (H{\alp}), and they verified that there is a
prominent population of candidate PMS stars located in this part of the
CMD, ``that have likely formed together with NGC 346
about 3-5 Myr ago''.

In order to make a first-order characterization of the observed stellar 
species we select four regions in the CMD, where each corresponds to one 
of the most prominent observed CMD features (UMS, LMS, RGB and PMS). The 
limits of the selected CMD regions are shown in Figure \ref{fig-cmd}. We 
distinguish, thus, four selected stellar groups, and we plot the spatial 
distribution of each one with the use of star counts.

\begin{figure*}[t!]
\centerline{\hbox{
\includegraphics[width=0.4685\textwidth,angle=0]{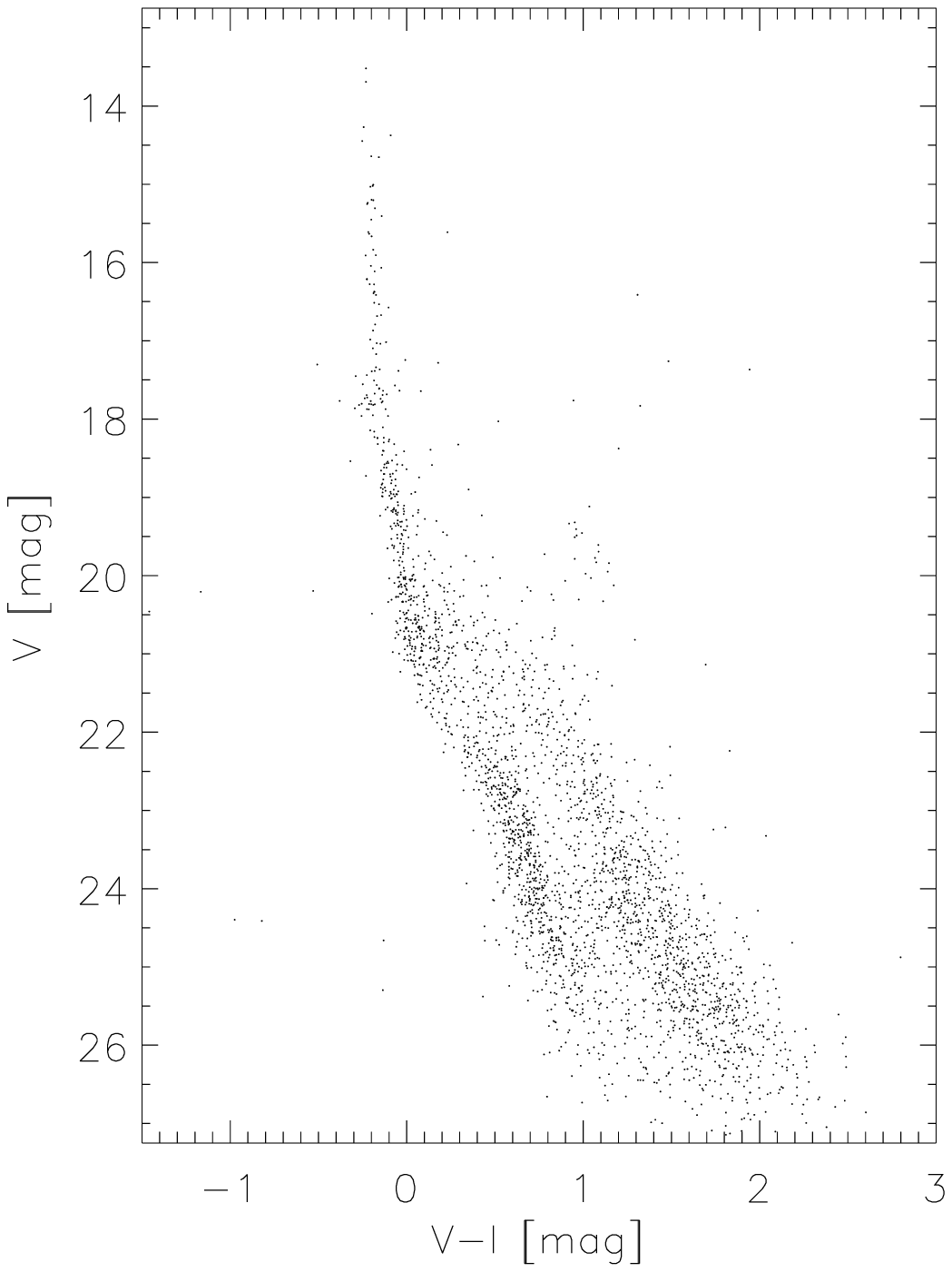}
\includegraphics[width=0.4685\textwidth,angle=0]{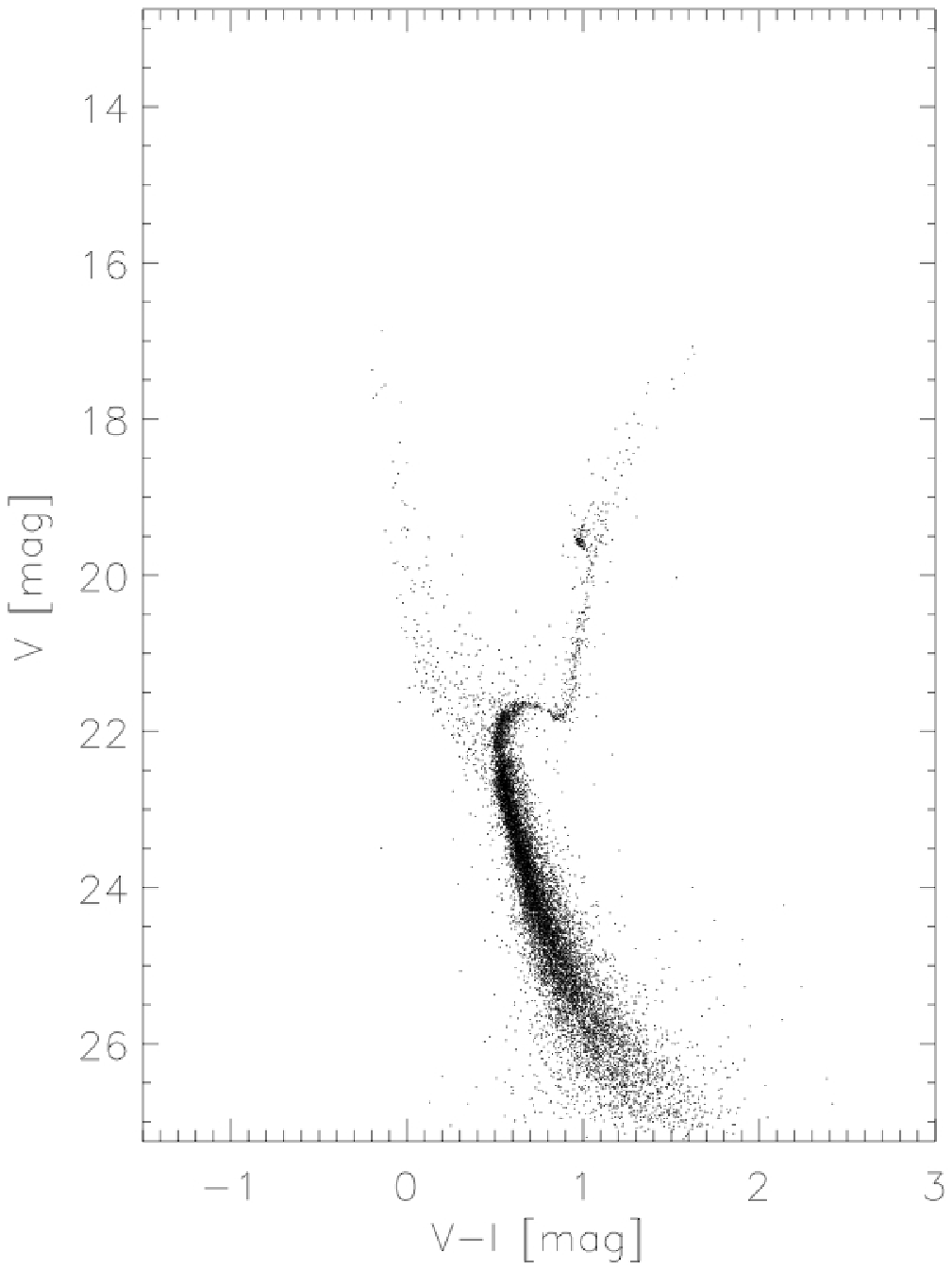}
}}
\caption{The $V-I$, $V$ CMDs of the stars from our catalog confined within 
a boxed 14.1 $\times$ 10.2 pc area centered on the main body of 
the stellar association NGC 346 (left panel) and a circular area of radius 
8.5 pc centered on the old star cluster BS 90 (right panel). The 
comparison of these CMDs exhibits the differences in stellar content of 
two stellar systems coexisting in the same region at a projected distance 
not larger than 24 pc from each other.} 
\label{fig-cmdsys}
\end{figure*}


\subsection{Characterization of the Observed Stellar Populations}

We identify the area in the observed region where each type of stars is 
dominant by performing star counts, and we construct the spatial 
density distribution of each stellar kind. We use the photometric 
catalog of the detected stars to perform star counts on square grids, 
under the assumption that each star is a point determined by its 
coordinates in the catalog. Each grid element has dimensions $\simeq$ 70 
WFC pixels, or $\simeq$ 3\farcs5, which corresponds to about 1 pc at the 
distance of the SMC of about 60 kpc (Dolphin et al. 2001). This star count 
algorithm is originally used for the detection of stellar associations and 
open clusters in the Large Magellanic Cloud by Gouliermis et al. (2000), 
who also give a detailed description of the method.

The high resolution of ACS/WFC imaging and the large stellar sample allow 
the construction of detailed maps of the spatial density distributions for 
each stellar group (isodensity contour maps). The constructed contour maps 
are shown in Figure \ref{fig-con}. The lowest isopleth corresponds to the 
background surface density and isopleths of higher surface density are 
plotted in steps of 1$\sigma$, where $\sigma$ is the standard deviation 
of the background density. Any stellar concentration with density equal to or 
higher than 3$\sigma$ above the background density (fourth isopleth in 
the maps of Figure \ref{fig-con}) is considered to be statistically 
significant. These maps show that not only different stellar populations 
dominate different areas of the observed region, but also their spatial 
distribution is very clumpy. The RGB and LMS groups seem to characterize 
mostly the massive old star cluster BS 90 (Bica \& Schmitt 1995), while the 
UMS and PMS stars are mostly concentrated in the general area of the 
association NGC 346. It is interesting to note that two compact 
concentrations of UMS and PMS stars can be seen in the corresponding maps 
to the north-east of the main association.

\section{The Dominant Stellar Concentrations}

The two dominant stellar systems in the area can be identified in the 
contour maps of Figure \ref{fig-con}, from which an indication of the 
extent of the systems can be easily derived. Specifically, we select the 
areas, centered on the association NGC 346 and the cluster BS 90, as they 
are defined by the 3$\sigma$ isopleths in Figure \ref{fig-con}. For BS 90 
we use the RGB and LMS maps and for the association the UMS and PMS ones. 
A circular area has been selected for the cluster, which has almost the 
same dimensions in both LMS and RGB contour maps. For the association we 
select a quadrilateral area, which confines the 3$\sigma$ isopleth in 
both PMS and UMS contour maps. This kind of selection is made because of 
the irregular, non-spherical shape of the association, as it is typical 
for associations, which are characterized by their extended form. The 
location of the star cluster BS 90 is well defined by a circular area with 
radius around 8.5 pc, and the one of the main body of the association NGC 
346 by a box of dimensions 14.1 $\times$ 10.2 pc.

Obviously the majority of the stars confined in each of the selected 
regions belongs to the populations of the corresponding systems and hence 
they represent the main stellar content of the systems. Consequently the 
CMDs of the stars in each region correspond to each of the systems. These 
CMDs plotted for stars in the quadrilateral area centered on NGC 346 and a 
circle centered on BS 90 are both shown in Figure \ref{fig-cmdsys}. These 
plots provide a first order distinction between the two dominant stellar 
concentrations in the star forming region of NGC 346/N 66, based on their 
stellar content. Consequently, these CMDs exhibit the variety of stellar 
species that are included in the relatively small observed 5\arcmin\ 
$\times$ 5{\arcmin} region. Although BS 90 is a very bound and obviously 
massive star cluster (see map of Figure \ref{fig-map}) it was never given 
much attention because of the fact that its stellar content does not 
include any blue massive stars, which were the main targets of previous 
works, and probably also because it was out-shined by the large number of 
bright blue super-giants of the association (see introduction).

Naturally, each of the CMDs shown in Figure \ref{fig-cmdsys} does not 
fully account only for the stellar population in each of the corresponding 
systems, because it is contaminated by the contribution not only of the 
general field of SMC at this region, which deserves an equally thorough 
study, but also of the population of the other system. This is more 
obvious in the CMD of BS 90 (right panel), where the upper main sequence 
is populated by a small number of blue giants up to $V \simeq 17$ mag. 
Probably this contamination comes from the association NGC 346 itself. The 
CMD of BS 90 is extremely sharp with a very clear turn-off and red giants 
branch, that make it a template cluster for the study of old clustered 
populations in the low-metallicity SMC. A more detailed study on this 
cluster and its environment is under way for the investigation of the 
fundamental properties of the stellar content and the dynamical behavior 
of this cluster (Rochau et al. in preparation).

On the other hand the CMD of the area centered on NGC 346 association
seems to be more clear from any old population contamination. A very
interesting feature of this CMD is the very sharp upper main sequence,
which is easily fitted by the ZAMS, making the age determination of the
association from these data alone very difficult. Another interesting
feature in this CMD is the secondary faint red sequence almost parallel
to the lower main sequence, which obviously corresponds to the PMS
population of the association in agreement with the findings of Nota et
al. (2006).  Furthermore, the higher detection limit in comparison to BS
90 is a clear indication of the higher extinction that takes place in
the area of the association. The case of the association NGC 346 is
currently under detailed study by us for the specification of the extent
of this system, the contamination of its CMD from the general field and
the determination of its massive and low-mass Initial Mass Function
(Hennekemper et al. in preparation).

\section{Final Remarks}

We take advantage of the large improvement in sensitivity and wide-field 
resolution provided by ACS to perform a detailed photometric study of the 
most active star forming region in the SMC. We use the ACS module of the 
photometric package DOLPHOT, which is especially designed for imaging with 
the ACS and we provide the full photometric catalog of all stars detected 
with short and long exposures in three ACS/WFC fields, that covers a 
region of about 5\arcmin\ $\times$ 5\arcmin\ centered on the young stellar 
association NGC 346. This region is a typical example of mixed stellar 
populations in the low-metallicity environment of the SMC. The detailed 
study of the stellar systems and the ambient general field of SMC in the 
region will provide new constraints for the determination of fundamental 
properties of concentrated stellar populations, such as relaxation 
time-scales and their Mass Functions.

We distinguish on the observed CMD of almost 100,000 detected stars in
total, four regions, as the most representative of different varieties:
Upper Main Sequence (UMS), Red Giant Branch (RGB), Lower Main Sequence
(LMS) and Pre-Main Sequence (PMS) stars. We find that UMS and PMS
represent mostly one of the two dominant stellar concentrations in the
observed region, the stellar association NGC 346, while RBG and LMS are
mostly populated at the recently discovered old star cluster BS 90 (Bica
\& Schmitt 1995). The CMDs of the stars confined in the main central
part of each of these two systems exhibit their differences in terms of
their stellar content: The association NGC 346 is a well extended star
forming system, while the cluster BS 90 is an old massive spherical star
cluster, which can be considered a template for old low-metallicity
populations.

Two investigations, one on BS 90 (Rochau et al. in preparation) and the
other on the association NGC 346 (Hennekemper et al. in preparation) are
under way to throw more light on these completely different neighboring
stellar concentrations in the SMC. These systems are only a part of the
variety of interesting objects, that are located in the observed region.
Such are the components of the {\sc Hii} region N 66 (Henize 1956), as
they are distinguished by Davies et al. (1976), small young star
clusters (Oey et al. 2004), and CO clouds (Rubio et al. 2000). In
addition, the region is characterized by a plethora of X-ray sources
(Naz{\'e} et al. 2003), emission stars and small nebulae (Meyssonnier \&
Azzopardi 1993), and radio sources (Loiseau et al.  1987; Filipovic et
al. 2002), as well as luminous IR sources, revealed with ISO and IRAS
(Wilke et al. 2003) and {\em Spitzer} (Brandl, 2005). Therefore we make
the stellar catalog resulted from our photometry available, so that the
whole astronomical community can benefit from these extremely useful and
accurate data.

\acknowledgments

D. Gouliermis acknowledges the support of the German Research Foundation 
(Deutsche Forschungsgemeinschaft - DFG) through the individual grant 
1659/1-1. This paper is based on observations made with the NASA/ESA 
Hubble Space Telescope, obtained from the data archive at the Space 
Telescope Science Institute. STScI is operated by the Association of 
Universities for Research in Astronomy, Inc. under NASA contract NAS 
5-26555. This research has made use of NASA's Astrophysics Data System, 
{\em Aladin} (Bonnarel et al. 2000), {\em WCSTools} (Mink 2001), and the 
SIMBAD database, operated at CDS, Strasbourg, France.



\end{document}